\documentclass[aip,jmp,amsmath,amssymb,preprint,]{revtex4-1}
\usepackage{amssymb,amsmath,bm}
\usepackage{longtable}
\usepackage{multirow}
\usepackage{setspace}
\usepackage{booktabs}
\usepackage{hyperref}
\usepackage{cleveref}
\usepackage{color}
\usepackage{xcolor}
\usepackage{mathrsfs,amsmath}
\usepackage[version=4]{mhchem}
\usepackage{mathrsfs}
\usepackage{graphicx}
\usepackage{subfigure} 
\usepackage{dcolumn}
\usepackage{bm}
\usepackage{amsthm}
\usepackage{ulem}
\usepackage{xr}
\usepackage{setspace}
\Crefname{figure}{Fig.}{Figs.}
\Crefname{equation}{Eq.}{Eqs.}

\begin{document}
\newtheorem{thm}{Theorem}[section]
\newtheorem{lemma}[thm]{Lemma}
\newtheorem{prop}[thm]{Proposition}
\newtheorem{rem}[thm]{Remark}
\newtheorem{cor}[thm]{Corollary}

\title{Thermodynamics for Reduced Models of Breakable Amyloid Filaments Based on Maximum Entropy Principle}
\author{Xinyu Zhang}
\affiliation{School of Mathematics, Sun Yat-Sen University, Guangzhou, Guangdong, 510275, P. R. China}
\author{Haiyang Jia}
\affiliation{College of Mathematics and Data Science, Minjiang University, Fuzhou, Fujian, 350108, P. R. China}
\affiliation{School of Mathematics and Statistics, Fuzhou University, Fuzhou, Fujian, 350108, P. R. China}
\author{Wuyue Yang}
\affiliation{Beijing Institute of Mathematical Sciences and Applications, Beijing, 101408, P. R. China}
\author{Liangrong Peng}
\email{peng@mju.edu.cn}
\affiliation{College of Mathematics and Data Science, Minjiang University, Fuzhou, Fujian, 350108, P. R. China}
\author{Liu Hong}
\email{hongliu@sysu.edu.cn}
\affiliation{School of Mathematics, Sun Yat-Sen University, Guangzhou, Guangdong, 510275, P. R. China}

\begin{abstract}
Amyloid filaments are associated with neurodegenerative diseases such as Alzheimer's and Parkinson's. Simplified models of amyloid aggregation are crucial because the original mass-action equations involve numerous variables, complicating analysis and understanding. While dynamical aspects of simplified models have been widely studied, their thermodynamic properties are less understood. In this study, we explore the Maximum Entropy Principle (MEP)-reduced models, initially developed for dynamical analysis, from a brand-new thermodynamic perspective. Analytical expressions along with numerical simulations demonstrate that the discrete MEP-reduced model strictly retains laws of thermodynamics, which holds true even when filament lengths transit from discrete values 
to continuous real numbers. Our findings not only clarify the thermodynamic consistency between the MEP-reduced models and the original models of amyloid filaments for the first time, but also suggest avenues for future research into the model-reduction thermodynamics.
\end{abstract}


\maketitle

\section{Introduction}
The Maximum Entropy Principle (MEP), originally proposed by Jaynes\cite{jaynes1985macroscopic}, offers a robust framework for statistical inference. In order to deduce unknown information from limited knowledge, its basic idea is to find the one (usually the probability or probability density) which maximizes the system entropy subject to pre-given constraints. In such a way, the MEP fits well into the framework of statistical physics, and has been widely applied to diverse fields such as statistical physics\cite{Topsoe2007}, biology\cite{kleidon2004non}, ecology\cite{harte2011maximum}, pure mathematics\cite{JMP2002,JMP2018} and emerging technologies like deep reinforcement learning\cite{haarnoja2018soft}. Later, the MEP has been extended to the Maximum Caliber principle (MCP), and used for deducing the most probable dynamical trajectories based on limited data \cite{Presse2013Principles, KenADill2012, KenADill2019}. 

Restricted to the subject of model reduction, the MEP also plays a significant role. For example, Anwasia et al. \cite{Anwasia2022} applied the MEP to derive an approximate velocity distribution function for Boltzmann equations in mixtures, resulting in a non-isothermal Maxwell-Stefan diffusion model. For high-dimensional chemical master equations describing complex chemical reactions, the application of the MEP allowed to express the unknown probability distribution solution through several simple macroscopic moments, thus enabling the closure of moment equations and a dramatic reduction of coupled ODE systems simultaneously\cite{Hong2013SimpleMM}. Karlin \cite{Karlin2016} applied the MEP to derive a reduced Fokker-Planck equation with closed moments, providing explicit formulas for the lowest eigenvalue and corresponding eigenfunction. Horvat et al. \cite{Horvat2015} applied the MEP to address the degeneracy problem in network models by transforming the density of states function to be log-concave. Esen et al. \cite{JMP2022,JMP2024} established the equivalence between the MEP and Onsager's variational principle within the framework of GENERIC (General Equation for
the Non-Equilibrium Reversible-Irreversible Coupling). 
 
In the above studies, researchers focused primarily on the accuracy of dynamical solutions of the reduced model compared to those of the original model. However, the thermodynamic consistency of models before and after applying reducing methods is also crucial. For instance, it has been shown that with respect to closed chemical reaction networks, the reduced models by either Partial Equilibrium Approximation (PEA) or Quasi-Steady-State Approximation (QSSA) may not preserve the thermodynamic structure of the original full model, particularly when algebraic relations are used instead of differential equations \cite{Hong2023}. A similar investigation has been carried out by Esposito et al.\cite{Esposito2020}, who applied the QSSA to simplify an open chemical reaction network and constructed the thermodynamics for the corresponding reduced models. 

We notice such kind of thermodynamic consistency check on the MEP is still lacking, which constitutes the basic motivation of the current study. Furthermore, once the thermodynamic consistency is provided, a nice thermodynamic structure, including several key thermodynamics quantities and their relations, for the reduced models by the MEP could be established accordingly. This is nearly impossible by directly considering the reduced models and without referring to the knowledge about the full model.  


To provide a concrete illustration, here we consider the dynamical models for amyloid filaments formation. Amyloid filaments formation is a ubiquitous biochemical phenomenon, which describes the transition procedure of amyloid proteins from soluble monomeric structures to insoluble fibrillar structures \cite{Chiti2006,Molecules25051195,Louros2023}. Its physiological significance has been fully recognized in the past years as the primary triggers for severe neurodegenerative diseases such as Alzheimer's disease, Parkinson's disease, and Type II diabetes \cite{Spillantini1997, Selkoe2001, Aguzzi2010, Westermark2011,LC2022}. The formation of amyloid filaments involves a series of complex chemical reactions, which can be described in a quantitative way by utilizing kinetic models in the form of differential equations\cite{OOSAWA196210,Lomakin1997,Knowles2009Science}. In this way, the underlying molecular mechanisms of amyloid filaments formation has been fully revealed. 

In an early publication\cite{Hong2013SimpleMM}, we adopted the MEP to simplify the complicated dynamical models for amyloid filaments formation, especially in the presence of length-dependent fragmentation processes. The applicability of the MEP in this case has been well justified through extensive numerical comparisons, though all restricted to the dynamical aspect. Therefore, in this paper we want to explore this issue further from a totally different thermodynamic view. To be concrete, the chemical kinetics and thermodynamics of the full model for the formation of breakable amyloid filaments, as well as the dynamics of the reduced models based on the MEP are introduced in \cref{sec:Models and method}. The thermodynamics of the MEP-reduced models are examined with great care in \cref{sec:Results}, including both analytical results and numerical simulations. The conclusion and discussion are presented in \cref{sec:Conclusion}. 


\section{Kinetic Models and Their Reduction By MEP}
\label{sec:Models and method} 
The following content is divided according to whether the state variable is discrete or continuous and whether the reaction mechanism contains the fragmentation process or not. According to the general theory of chemical reactions, a discrete dynamical model for amyloid filaments formation is presented in \cref{subsubsec:dm}, whose continuous limit is given in \cref{subsubsec:cl}. The corresponding reduced models by applying the principle of maximum entropy are shown in \cref{subsubsec:rdm} and \cref{subsubsec:rcm} respectively. Besides dynamical models, preliminary results on the thermodynamics of the full models are presented too.

\subsection{Mathematical Models for Amyloid filaments formation}
\begin{figure}
    \centering
    \includegraphics[width=0.75\linewidth]{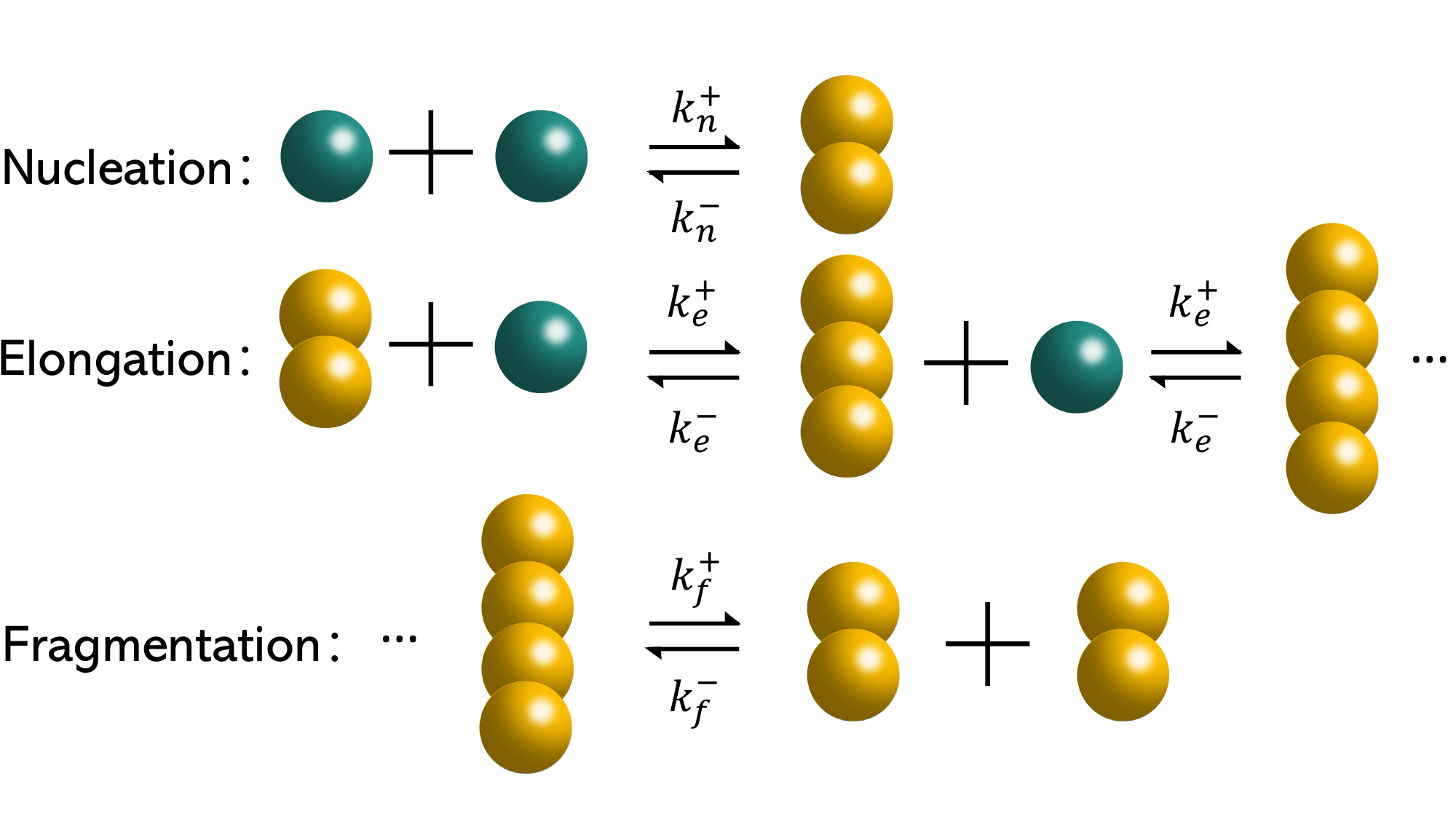}
    \caption{Illustration of primary nucleation, elongation, fragmentation and their reverse processes during the formation of breakable amyloid filaments. Here the blue balls denote monomeric proteins, yellow balls denote filaments.}
    \label{fig:ball}
\end{figure}

Amyloid proteins or peptides can form highly ordered, insoluble aggregates, that are resistant to enzyme degradation, in a self-assembling way. To quantitatively account for this formation procedure, a widely adopted model includes three basic processes–
primary nucleation, elongation, and fragmentation\cite{Hong2017,Knowles2015,Dobson2014}, \textit{i.e.},
\begin{equation}
\label{CRN}
    \begin{array}{l}
n_{c} A_{1} \underset{k_{n}^{-}}{\stackrel{k_{n}^{+}}{\rightleftharpoons}} A_{n_{c}}, \\
A_{1}+A_{i} \underset{k_{e.}^{-}}{\stackrel{k_{e}^{+}}{\rightleftharpoons}} A_{i+1}, \quad\left(i \geq n_{c}\right)\\
A_{i+j} \underset{k_{f}^{-}}{\stackrel{k_{f}^{+}}{\rightleftharpoons}} A_{i}+A_{j}, \quad\left(i,j \geq n_{c}\right)
\end{array}
\end{equation}
where $A_1$ denotes monomeric amyloid proteins, $A_i$ denotes filaments of size $i$ (meaning constituted by $i$ monomeric proteins). The number $n_c$ gives the critical nucleus size ($n_c\geq 2$), while the constants $k_{n}^{+}, k_{e}^{+}, k_{f}^{+}$ (resp. $k_{n}^{-}, k_{e}^{-}, k_{f}^{-}$) are the forward (resp. backward) reaction rate constants for fibril nucleation, elongation and fragmentation, respectively. See Fig. \ref{fig:ball} for an illustration. As a special case of the full model in \eqref{CRN}, we will also discuss scenarios where the filament fragmentation reactions can be neglected, meaning that the model consists only of nucleation and elongation reactions. Furthermore, the model in \eqref{CRN} is considered to be a closed chemical reaction network (CRN), since no species is allowed to exchange with the external environment. 

In the next part, we will discuss the Discrete model and its Continuous limit separately.

\subsubsection{Discrete model}
\label{subsubsec:dm}
Let us take $[A_1](t)$ as the molar concentration of monomeric proteins, and $\{[A_i](t)_{i=n_c}^\infty\}$ as the concentration distribution of filaments of various lengths. Then referring to the molecular mechanism illustrated in \eqref{CRN} and laws of mass-action, the time-evolution equations for the concentrations of aggregates \(([A_1], [A_i]_{i \geq n_c})\) can be written as 
\begin{equation}
\label{dAdt}
\begin{cases}
&\frac{d}{dt}\left[A_{1}\right]
     =-n_{c}{k}_{n}^{+}[{A_{1}}]^{{n_{c}}}+n_{c}{k}_{n}^{-}[A_{{n_{c}}}]-2{k}_{e}^{+}\left[A_{1}\right]{\sum }_{j=n_{c}}^{\mathrm{\infty }}\left[A_{j}\right]+2{k}_{e}^{-}{\sum }_{j=n_{c}+1}^{\mathrm{\infty }}\left[A_{j}\right], \\
&\frac{d}{dt}\left[A_{i}\right]
 =2{k}_{e}^{+}\left[A_{1}\right]\left(\left[A_{i-1}\right]-\left[A_{i}\right]\right)-2{k}_{e}^{-}\left(\left[A_{i}\right]-\left[A_{i+1}\right]\right)\\
& +{\sum }_{j=n_{c}+i}^{\mathrm{\infty }}2{k}_{f}^{+}\left[A_{j}\right]-{\sum }_{j=n_{c}}^{i-n_{c}}{k}_{f}^{+}\left[A_{i}\right]-{\sum }_{j=n_{c}}^{\mathrm{\infty }}2{k}_{f}^{-}\left[A_{i}\right]\left[A_{j}\right]+{\sum }_{j=n_{c}}^{i-n_{c}}{k}_{f}^{-}\left[A_{j}\right]\left[A_{i-j}\right]\\
& +({k}_{n}^{+}[{A_{1}}]^{{n_{c}}}-{k}_{n}^{-}[A_{i}]-2{k}_{e}^{+}[A_{1}][A_{i-1}]+2{k}_{e}^{-}[A_{i}])\delta _{i,{n_{c}}},~(i\geq n_{c}) .
\end{cases}
\end{equation}
The initial values are set as \(([A_1],[A_i]_{i \geq n_c})|_{t=0} = (m_{tot}, 0,\cdots,0)\). The coefficient 2 in front of $k_e^{\pm}$ represents that the reactions can proceed independently at both ends of a filament. Here the conservation law of amyloid proteins, $m_{tot}=\left[A_{1}\right]+{\sum }_{i=n_{c}}^{\mathrm{\infty }}i\left[A_{i}\right]$, is readily guaranteed, which is proved in SI. The chemical mass-action equations in \eqref{dAdt} is termed as the \emph{Discrete model}. 

A steady state of the above equations is reached when the production and consumption rates of each substance become equal: 
\begin{equation}
    \frac{d[A_i]^{ss}}{dt}=0,\quad i=1,n_c,n_c+1,\cdots.
    \label{ness}
\end{equation}
There exists a special steady state, known as the equilibrium state, which is defined to have zero net flux (the forward flux minus the backward flux) for every reactions. It can be shown that the equilibrium state $\{ \left[A_{i}^e \right] \}$ is attained if and only if the respective forward and backward reaction rate functions equal to each other, \textit{i.e.}, 
\begin{equation}
\begin{array}{rl}
&{k}_{n}^{+}(\left[A_{1}\right]^e)^{{n_{c}}}  ={k}_{n}^{-}\left[A_{{n_{c}}}\right]^e, \\
&{k}_{e}^{+}\left[A_{1}\right]^e\left[A_{i}\right]^e  ={k}_{e}^{-}\left[A_{i+1}\right]^e,\quad i=n_{c},n_{c}+1,\ldots, \\
&{k}_{f}^{+}\left[A_{i+j}\right]^e  ={k}_{f}^{-}\left[A_{i}\right]^e\left[A_{j}\right]^e,\quad i,j=n_{c},n_{c}+1,\cdots. 
\end{array}
\end{equation}
This constraint on the reaction rate constants is called the condition of detailed balance. 

Now we proceed to the thermodynamics of fibril formation described in  \eqref{CRN} under the detailed balance condition. Firstly, the entropy function reads 
\begin{equation}
\begin{array}{rl}
\label{ent}
Ent\left(t\right)
&=-\left(\left[A_{1}\right]\ln \left[A_{1}\right]-\left[A_{1}\right]\right)-{\sum }_{i=n_{c}}^{\mathrm{\infty }}\left(\left[A_{i}\right]\ln \left[A_{i}\right]-\left[A_{i}\right]\right).
\end{array}
\end{equation}
According to the general theory of non-equilibrium thermodynamics\cite{deGroot1962,Hong2015}, its time derivative ${d Ent\left(t\right)}/{dt}$ can be separated into two different contributions -- the rates of entropy flow $J_{f}\left(t\right)$ and entropy production $epr\left(t\right)$,
\begin{equation}
\begin{array}{rl}
J_{f}\left(t\right)=&-\left[\ln \frac{{k}_{n}^{+}}{{k}_{n}^{-}}\left(R^{n+}-R^{n-}\right)\right]-{\sum }_{i=n_{c}}^{\mathrm{\infty }}\left[\ln \frac{{k}_{e}^{+}}{{k}_{e}^{-}}\left({R}_{i}^{e+}-{R}_{i}^{e-}\right)\right]\\
& -{\sum }_{i=n_{c}}^{\mathrm{\infty }}\ln \frac{{k}_{f}^{+}}{{k}_{f}^{-}}{\sum }_{j=n_{c}}^{\mathrm{\infty }}\left({R}_{ij}^{f+}-{R}_{ij}^{f-}\right)+\frac{1}{2}{\sum }_{i=2n_{c}}^{\mathrm{\infty }}{\sum }_{j=n_{c}}^{\mathrm{i}-\mathrm{n}_{\mathrm{c}}}\ln \frac{{k}_{f}^{+}}{{k}_{f}^{-}}\left({R}_{\left(i-j\right),j}^{f+}-{R}_{\left(i-j\right),j}^{f-}\right),\\
epr\left(t\right)=&\left[\ln \frac{R^{n+}}{R^{n-}}\left(R^{n+}-R^{n-}\right)\right]+{\sum }_{i=n_{c}}^{\mathrm{\infty }}\left[\ln \frac{{R}_{i}^{e+}}{{R}_{i}^{e-}}\left({R}_{i}^{e+}-{R}_{i}^{e-}\right)\right]\\
&+{\sum }_{i=n_{c}}^{\mathrm{\infty }}{\sum }_{j=n_{c}}^{\mathrm{\infty }}\ln \frac{{R}_{ij}^{f+}}{{R}_{ij}^{f-}}\left({R}_{ij}^{f+}-{R}_{ij}^{f-}\right)-\frac{1}{2}{\sum }_{i=2n_{c}}^{\mathrm{\infty }}{\sum }_{j=n_{c}}^{\mathrm{i}-\mathrm{n}_{\mathrm{c}}}\ln \frac{{R}_{\left(i-j\right),j}^{f+}}{{R}_{\left(i-j\right),j}^{f-}}\left({R}_{\left(i-j\right),j}^{f+}-{R}_{\left(i-j\right),j}^{f-}\right)\geq 0,
\end{array}
\end{equation}
where $R^{n+}={k}_{n}^{+}[{A_{1}}]^{{n_{c}}}$, $R_{i}^{e+}={k}_{e}^{+}[A_{1}][A_{i}]$ and ${R}_{ij}^{f+}={k}_{f}^{+}[A_{i+j}]$ denote the reaction rate functions for primary nucleation, elongation and fragmentation separately. So are $R^{n-}={k}_{n}^{-}[A_{{n_{c}}}]$, $R_{i}^{e-}={k}_{e}^{-}[A_{i+1}]$ and ${R}_{ij}^{f-}={k}_{f}^{-}[A_{i}][A_{j}]$ for those of the reverse processes. The entropy production rate $epr(t)$ characterizes the system dissipation and provides a natural way to measure the distance from the equilibrium state, which is always non-negative in accordance with the second law of thermodynamics; while the entropy flow rate $J_{f}\left(t\right)$ quantifies the amount of heat exchanged with the environment, which has an indefinite sign.

Besides the entropy, the free energy function could also be introduced, that is
\begin{equation}
\label{F_original}
F(t)
=\left([A_{1}]\ln \frac{[A_{1}]}{[A_{1}]^e}-[A_{1}]+[A_{1}]^e \right)+{\sum }_{i=n_{c}}^{\mathrm{\infty }}\left([A_{i}]\ln \frac{[A_{i}]}{[A_{i}]^e}-[A_{i}]+[A_{i}]^e \right), 
\end{equation}
which serves as a Lyapunov function for the \emph{Discrete model}  in \eqref{dAdt}, by satisfying $F(t) \geq 0$ and $dF(t)/dt \leq 0$. The free energy dissipation rate is defined as
\begin{equation}
    f_{d}\left(t\right)  \equiv -\frac{dF(t)}{dt}=\left(\frac{d[A_{1}]}{dt}\ln \frac{[A_{1}]}{[A_{1}]^e} \right)+{\sum }_{i=n_{c}}^{\mathrm{\infty }}\left(\frac{d[A_{i}]}{dt}\ln \frac{[A_{i}]}{[A_{i}]^e}\right)\geq 0,
\end{equation}
by recalling that the term $[A_i]^e$ represents the concentration of 
$A_i$ at the equilibrium. 

\begin{rem}
For CRNs occurring within a closed system, the equivalence of the free energy dissipation rate and the entropy production rate, denoted as ${f_{d}}\left(t\right) = epr(t)$,  is established exclusively under the condition of detailed balance. Otherwise, when this condition is broken, such as by chemostatting some species, one can consider their difference, $Q_{hk}(t)=epr(t)-f_{d}(t)$, where $Q_{hk}(t)$ represents the housekeeping entropy production rate \cite{Sekimoto2010, Sasa2022, Esposito2020, Qian2017}. Notably, $Q_{hk}(t)$ is invariably non-negative and attains its minimum (zero value) provided the condition of complex balance holds in open CRNs. 
\end{rem}

\subsubsection{Continuous limit}
\label{subsubsec:cl}
By assuming the length of each unit element of a filament as infinitesimally small, we can take the continuous limit and replace the differences in \eqref{CRN} by differentials and the summations by integrals. This approach allows us to arrive at the following differential-integral equations:
\begin{equation}
\label{continuous model}
\begin{cases} 
&\frac{d}{dt}\left[A_{1}\right]  =-n_{c}{k}_{n}^{+}[{A_{1}}]^{{n_{c}}}+n_{c}{k}_{n}^{-}\left[A_{{n_{c}}}\right]-2{k}_{e}^{+}\left[A_{1}\right]{\int }_{n_{c}}^{\infty }\left[A_{y}\right]dy+2{k}_{e}^{-}{\int }_{n_{c}}^{\infty }\left[A_{y}\right]dy, \\
&\frac{\partial }{\partial t}\left[A_{x}\right]
 =-2{k}_{e}^{+}\left[A_{1}\right]\frac{\partial }{\partial x}\left[A_{x}\right]+2{k}_{e}^{-}\frac{\partial }{\partial x}\left[A_{x}\right]\\
& +2{k}_{f}^{+}{\int }_{n_{c}+x}^{\infty }\left[A_{y}\right]dy-2{k}_{f}^{-}\left[A_{x}\right]{\int }_{n_{c}}^{\infty }\left[A_{y}\right]dy-{k}_{f}^{+}{\int }_{n_{c}}^{x-n_{c}}\left[A_{x}\right]dy+{k}_{f}^{-}{\int }_{n_{c}}^{x-n_{c}}\left[A_{y}\right]\left[A_{x-y}\right]dy\\
& +\left({k}_{n}^{+}\left[A_{1}\right]^{{n_{c}}}-{k}_{n}^{-}\left[A_{{n_{c}}}\right]-2{k}_{e}^{+}\left[A_{1}\right]\left[A_{{n_{c}}}\right]+2{k}_{e}^{-}\left[A_{{n_{c}}}\right]\right)\delta _{x,{n_{c}}} ~\left(x\geq n_{c}\right), 
\end{cases}
\end{equation}
in which $[A_1]$ and $[A_x]$ represent the molar concentration of monomeric proteins and filaments of length $x$ respectively. Now the index $x$ is a real number instead of discrete integer. 
Note that we adopt the constraint that the integral \(\int_a^b f(y) \, dy\) becomes zero when \(a \geq b\). This guarantees the consistency with the discrete model when this term appears in the above  continuous equations. The initial condition is taken as $[A_1](0)=m_{tot}$ and $[A_x](0)=0$ for $\forall x\geq n_c$, which is accompanied by the infinite boundary condition, $\lim_{x\to \infty}[A_x](t)=0$. Eq. \eqref{continuous model} is termed as the \emph{Continuous model}. 

The condition of detailed balance of  the continuous model is outlined as 
\begin{equation}
\begin{array}{rl}
&{k}_{n}^{+}(\left[A_{1}\right]^e)^{{n_{c}}}  ={k}_{n}^{-}\left[A_{{n_{c}}}\right]^e, \\
&{k}_{e}^{+}\left[A_{1}\right]^e\left[A_{x}\right]^e  ={k}_{e}^{-}\left[A_{x}\right]^e,\quad x>n_{c}, \\
&{k}_{f}^{+}\left[A_{x+y}\right]^e  ={k}_{f}^{-}\left[A_{x}\right]^e\left[A_{y}\right]^e,\quad x,y>n_{c}. 
\end{array}
\end{equation}

Referring to previous thermodynamic quantities for the discrete model, the corresponding entropy function for the continuous model in  Eq. \eqref{continuous model} is given by
\begin{equation}
\label{entropy_cont}
    Ent\left(t\right) = -\left(\left[A_{1}\right]\ln \left[A_{1}\right]-\left[A_{1}\right]\right)-{\int }_{n_{c}}^{\infty }(\left[A_{y}\right]\ln \left[A_{y}\right]-\left[A_{y}\right])dy.
\end{equation}
Similarly, its time derivative can be separated into the rates of entropy flow $J_{f}\left(t\right)$ and entropy production $epr\left(t\right)$, i.e.
\begin{equation}
\begin{array}{rl}
&J_{f}\left(t\right)=-\left[\ln \frac{{k}_{n}^{+}}{{k}_{n}^{-}}\left({R}^{n+}-{R}^{n-}\right)\right]-{\int }_{n_{c}}^{\mathrm{\infty }}\left[\ln \frac{{k}_{e}^{+}}{{k}_{e}^{-}}\left({R}_{x}^{e+}-{R}_{x}^{e-}\right)\right]dx\\
& -{\int }_{n_{c}}^{\mathrm{\infty }}\ln \frac{{k}_{f}^{+}}{{k}_{f}^{-}}{\int }_{n_{c}}^{\mathrm{\infty }}\left({R}_{xy}^{f+}-{R}_{xy}^{f-}\right) dxdy
+\frac{1}{2}{\int }_{2n_{c}}^{\mathrm{\infty }}{\int }_{n_{c}}^{\mathrm{x}-\mathrm{n}_{\mathrm{c}}}\ln \frac{{k}_{f}^{+}}{{k}_{f}^{-}}\left({R}_{\left(x-y\right),y}^{f+}-{R}_{\left(x-y\right),y}^{f-}\right)dydx,\\
&epr\left(t\right)=\left[\ln \frac{{R}^{n+}}{{R}^{n-}}\left({R}^{n+}-{R}^{n-}\right)\right]+{\int}_{n_{c}}^{\mathrm{\infty }}\left[\ln \frac{{R}_{x}^{e+}}{{R}_{x}^{e-}}\left({R}_{x}^{e+}-{R}_{x}^{e-}\right)\right]dx\\
&+{\int }_{n_{c}}^{\mathrm{\infty }}{\int}_{n_{c}}^{\mathrm{\infty }}\ln \frac{{R}_{xy}^{f+}}{{R}_{xy}^{f-}}\left({R}_{xy}^{f+}-{R}_{xy}^{f-}\right)dxdy-\frac{1}{2}{\int }_{2n_{c}}^{\mathrm{\infty }}{\int}_{n_{c}}^{\mathrm{x}-\mathrm{n}_{\mathrm{c}}}\ln \frac{{R}_{\left(x-y\right),y}^{f+}}{{R}_{\left(x-y\right),y}^{f-}}\left({R}_{\left(x-y\right),y}^{f+}-{R}_{\left(x-y\right),y}^{f-}\right)dydx,
\end{array}
\end{equation}
where $R^{n+}={k}_{n}^{+}[{A_{1}}]^{{n_{c}}}, R^{n-}={k}_{n}^{-}[A_{{n_{c}}}]$, $R_{x}^{e+}={k}_{e}^{+}[A_{1}][A_{x}], R_{x}^{e-}={k}_{e}^{-}[A_{x}]$, and ${R}_{xy}^{f+}={k}_{f}^{+}[A_{x+y}], {R}_{xy}^{f-}={k}_{f}^{-}[A_{x}][A_{y}]$. The entropy production rate \(epr(t) \geq 0\) and entropy flow rate \(J_{f}(t)\) in the continuous model retain the nice properties outlined for the discrete model.

Besides the entropy, the free energy function in a continuous form reads 
\begin{equation}
\label{F_original_c}
F(t)
=\left([A_{1}]\ln \frac{[A_{1}]}{[A_{1}]^e}-[A_{1}]+[A_{1}]^e \right)+{\int}_{n_{c}}^{\mathrm{\infty }}\left([A_{x}]\ln \frac{[A_{x}]}{[A_{x}]^e}-[A_{x}]+[A_{x}]^e \right)dx, 
\end{equation}
which serves as a Lyapunov function for the continuous model \eqref{continuous model}, by satisfying $F(t) \geq 0$ and $dF(t)/dt \leq 0$. The free energy dissipation rate ${f_{d}}\left(t\right)  \equiv -dF(t)/dt$ becomes
\begin{equation}
    f_{d}=\left(\frac{\partial[A_{1}]}{\partial t}\ln \frac{[A_{1}]}{[A_{1}]^e} \right)+{\int }_{n_{c}}^{\mathrm{\infty }}\left(\frac{\partial[A_{x}]}{\partial t}\ln \frac{[A_{x}]}{[A_{i}]^e}\right)dx\geq 0,
\end{equation}
where the term $[A_x]^e$ represents the molar concentration of $A_x$  at the equilibrium for Eq. \eqref{continuous model}.

\subsection{Maximum Entropy Principle for Model Reduction}
The discrete model in \eqref{dAdt} is comprised by an infinite number of equations, while the continuous model in \eqref{continuous model} is an integral-differential equation, both of which are too complicated for applications. In addition, due to limitations on experimental instruments, information about macroscopic measurable quantities, like the fibril mass concentration, is much easier to be obtained than the filament length distribution $\{[A_i]\}_{i=n_c}^{\infty}$ (or $\{[A_x]\}_{x\geq n_c}$). Therefore, there is a desire to derive simplified models for macroscopic measurable quantities directly. The MEP provides a reliable way to perform such model reduction, and has been shown to be quite effective \cite{Hong2013SimpleMM}. Details of applying the MEP to our study on amyloid filaments formation are illustrated as follows. 

\subsubsection{Reduced discrete model}
\label{subsubsec:rdm}
In order to determine the optimal length distribution of filaments $\{[A_{i}]\}_{i=1, i\geq n_c}$ that maximizes the entropy function under pre-given constraints, we need to solve the following convex optimization problem: 
\begin{equation}
\begin{array}{l}
\max_{\{[A_{i}]\}_{i=1,n_{c}}^\infty} Ent\left(t\right),\\
\mathrm{s}.\mathrm{t}.\quad
{\sum }_{i=n_{c}}^{\mathrm{\infty }}\left[A_{i}\right]=P(t), \quad 
{\sum }_{i=n_{c}}^{\mathrm{\infty }}i \left[A_{i}\right]=M(t), \quad
\left[A_{1}\right]+{\sum }_{i=n_{c}}^{\mathrm{\infty }}i \left[A_{i}\right]=m_{tot},
\end{array}
\end{equation} 
where $P(t)$ and $M(t)$ represent the number and mass concentrations of filaments. And $m_{tot}$ is the total molar concentration of proteins, which is a constant due to the mass conservation law. By using the \emph{Lagrange multiplier method}, we can solve the above constrained optimization problem and obtain 
\begin{align*}
& \frac{\partial \mathcal{L}}{\partial \left[A_{1}\right]}=\lambda _{3}-\ln \left[A_{1}\right]=0,\\ 
& \frac{\partial \mathcal{L}}{\partial \left[A_{i}\right]}=\left[\lambda _{1}+i\left(\lambda _{2}+\lambda _{3}\right)\right]-\ln \left[A_{i}\right]=0,\quad i=n_{c},n_{c}+1,\ldots ,
\end{align*}
where the Lagrangian reads 
\begin{equation}
\begin{array}{rl}
\mathcal{L}(\{\left[A_{i}\right] \}; \{\lambda_{j}\}_{j=1,2,3} )=&Ent\left(t\right)+\lambda _{1}\left({\sum }_{i=n_{c}}^{\mathrm{\infty }}\left[A_{i}\right]-P\right)+\lambda _{2}\left( {\sum }_{i=n_{c}}^{\mathrm{\infty }}i \left[A_{i}\right]-M\right) \\
 &+\lambda _{3}\left( \left[A_{1}\right]+{\sum }_{i=n_{c}}^{\mathrm{\infty }}i \left[A_{i}\right]-m_{tot}\right).
 \end{array}
 \end{equation}
Substituting the solutions
\begin{equation*}
\left[A_{1}\right]=e^{{\lambda _{3}}}, \quad
\left[A_{i}\right]=e^{{\lambda _{1}}+i({\lambda _{2}}+{\lambda _{3}})},~i=n_{c},n_{c}+1,\ldots
\end{equation*}
into the constraints yields that
\begin{align*}
&\lambda_{1}  =\ln \frac{P^{2}}{M-(n_{c}-1)P}-n_{c}\ln \frac{M-n_{c}P}{M-(n_{c}-1)P}, \\
&\lambda_{2}  =\ln \frac{M-n_{c}P}{M-(n_{c}-1)P}-\ln \left(m_{tot}-M\right), \\
&\lambda_{3}  =\ln \left(m_{tot}-M\right).
\label{lambdad}
\end{align*}
Finally, the optimal length distribution of filaments is obtained as 
\begin{equation}
\label{optimal-dis}
\left[A_{1}\right]=m_{tot}-M, \quad
\left[A_{i}\right]=\frac{P}{L_d}\left(1-L_d^{-1}\right)^{i-n_c},~i=n_{c},n_{c}+1,\ldots,
\end{equation}
which are expressed as functions of macroscopic measurable quantities solely. An average length is defined as $L_d =[M - (n_c - 1)P]/{P}$.  

Now we are at a position to derive the reduced discrete model. Since $m_{tot}$ is a constant, we only need to consider the time evolution equations of $P(t)$ and $M(t)$. Combining the evolutionary equations for $[A_i]$ in Eq. \eqref{dAdt} and the optimal fibril length distribution in Eq. \eqref{optimal-dis}, we arrive at (see SI for a derivation)
\begin{equation}
\begin{cases} 
&\frac{dP}{dt}  =\hat{R}_n+\underline{\hat{R}_f},\\
&\frac{dM}{dt}  =n_c\hat{R}_n+2\hat{R}_e,
\end{cases} 
\label{Eq:kineticseqdy}
\end{equation}
where $\hat{R}_{\alpha} \equiv \hat{R}_{\alpha}^+-\hat{R}_{\alpha}^- (\alpha=n, e, f)$ and 
\begin{equation}
    \begin{array}{l}
    \hat{R}_n^+
    ={k}_{n}^{+}\left(m_{tot}-M\right)^{n_c}, \quad 
    \hat{R}_n^-={k}_{n}^{-}{PL_d^{-1}},\\
    \hat{R}_e^+
    ={k}_{e}^{+}\left(m_{tot}-M\right)P, \quad \hat{R}_e^-={k}_{e}^{-}P\left(1-L_d^{-1}\right),\\
    \hat{R}_f^+
    ={k}_{f}^{+}\left[M-\left(n_{c}-1\right)P\right]{\left(1-L_d^{-1}\right)}^{{n_{c}}}, \quad \hat{R}_f^-={k}_{f}^{-}P^{2}.
\end{array}
\label{fluxofdiscrete}
\end{equation}
The underlined term in Eq. \eqref{Eq:kineticseqdy} represents the contribution of fragmentation processes to the flux of $dP/dt$. Removing this term from Eq. \eqref{Eq:kineticseqdy} yields the reduced discrete model without fragmentation.

Provided the condition of detailed balance holds,  the equilibrium values of $P^e$ and $M^e$ can be expressed through \( m_{tot} \), \( {k_n^+}/{k_n^-} \) and  \( {k_e^+}/{k_e^-} \) as
\begin{equation}
   P^e=\frac{m_{tot}-[A_1]^e}{n_c-\frac{k_e^+}{k_e^-}[A_1]^e},\quad M^e=m_{tot}-[A_1]^e,
\label{dbc-discrete1}
\end{equation}
where $[A_1]^e$ is the solution of the following polynomial equation
\begin{equation}
    \left(\frac{k_n^+}{k_n^-}\right)\left(\frac{k_e^+}{k_e^-}\right)^{2}\left([A_1]^e\right)^{n_c+2}-\left(\frac{k_n^+}{k_n^-}\frac{k_e^+}{k_e^-}\right)(n_c+1)\left([A_1]^e\right)^{n_c+1}+n_c\left([A_1]^e\right)^{n_c}+[A_1]^e-m_{tot}=0.
\end{equation}

\subsubsection{Reduced continuous model}
\label{subsubsec:rcm}
In the continuous limit, we proceed to solve the following convex optimization problem 
\begin{equation}
\begin{array}{l}
\max _{[A_{x}]} Ent\left(t\right), \\
\mathrm{s}.\mathrm{t}.\quad
{\int }_{n_{c}}^{\infty }\left[A_{y}\right]dy=P, \quad  {\int }_{n_{c}}^{\infty }y\left[A_{y}\right]dy=M, \quad \left[A_{1}\right]+{\int }_{n_{c}}^{\infty }y\left[A_{y}\right]dy=m_{tot},
\end{array}
\end{equation}
where the entropy function is given in Eq. \eqref{entropy_cont}. Following the same procedures as for the discrete model, the optimal length distribution of filaments can be calculated in an explicit way as
\begin{equation}
\label{optimal-cont}
\left[A_{1}\right]=m_{tot}-M, \quad
\left[A_{x}\right]=\left(\frac{P}{L_c}\right)e^{-\frac{x-n_c}{L_c}},~x\geq n_{c},
\end{equation}
where $L_c =(M - n_c P)/{P}$ defines an effective average length. 

Now the time evolution equations for $P$ and $M$ in a continuous form read

\begin{equation}
\begin{cases} 
&\frac{dP}{dt}  =\check{R}_n+\underline{\check{R}_f},\\
&\frac{dM}{dt}  =n_c\check{R}_n+2\check{R}_e,
\end{cases} 
\label{Eq:kineticseqcy}
\end{equation}
where $\check{R}_{\alpha} \equiv \check{R}_{\alpha}^+-\check{R}_{\alpha}^-$ $(\alpha=n, e, f)$ and  
\begin{equation}
    \begin{array}{l}
    \check{R}_n^+
    ={k}_{n}^{+}\left(m_{tot}-M\right)^{n_c}, \quad 
    \check{R}_n^-={k}_{n}^{-}{PL_c^{-1}},\\
    \check{R}_e^+
    ={k}_{e}^{+}\left(m_{tot}-M\right)P, \quad \check{R}_e^-={k}_{e}^{-}P,\\
    \check{R}_f^+
    ={k}_{f}^{+}\left(M-n_{c}P\right)e^{-\frac{n_c}{L_c}}, \quad \check{R}_f^-={k}_{f}^{-}P^{2}.
    \end{array}
    \label{fluxofcontinuous}
    \end{equation}
The underlined term in Eq. \eqref{Eq:kineticseqcy} represents the contribution to flux from fragmentation. Removing this term from Eq. \eqref{Eq:kineticseqcy} straightforwardly yields the reduced continuous model without fragmentation.

Under the condition of detailed balance, $P^e$ and $M^e$ can be expressed through \( m_{tot} \), \({k_n^+}/{k_n^-} \) and \( {k_e^+}/{k_e^-} \) as 
\begin{equation}
\begin{array}{ll}
     M^e&=m_{tot}-[A_1]^e,\\
     2P^e&=-n_c\frac{k_n^+}{k_n^-}\left([A_1]^e\right)^{n_c}+\sqrt{\left[n_c\frac{k_n^+}{k_n^-}\left([A_1]^e\right)^{n_c}\right]^2+4\frac{k_n^+}{k_n^-}\left([A_1]^e\right)^{n_c}},
\end{array}
\label{dbc-continuous1}
\end{equation}
where $[A_1]^e=\frac{k_e^+}{k_e^-}$.

\begin{figure}[htbp]
    \centering
    \includegraphics[width=1\linewidth]{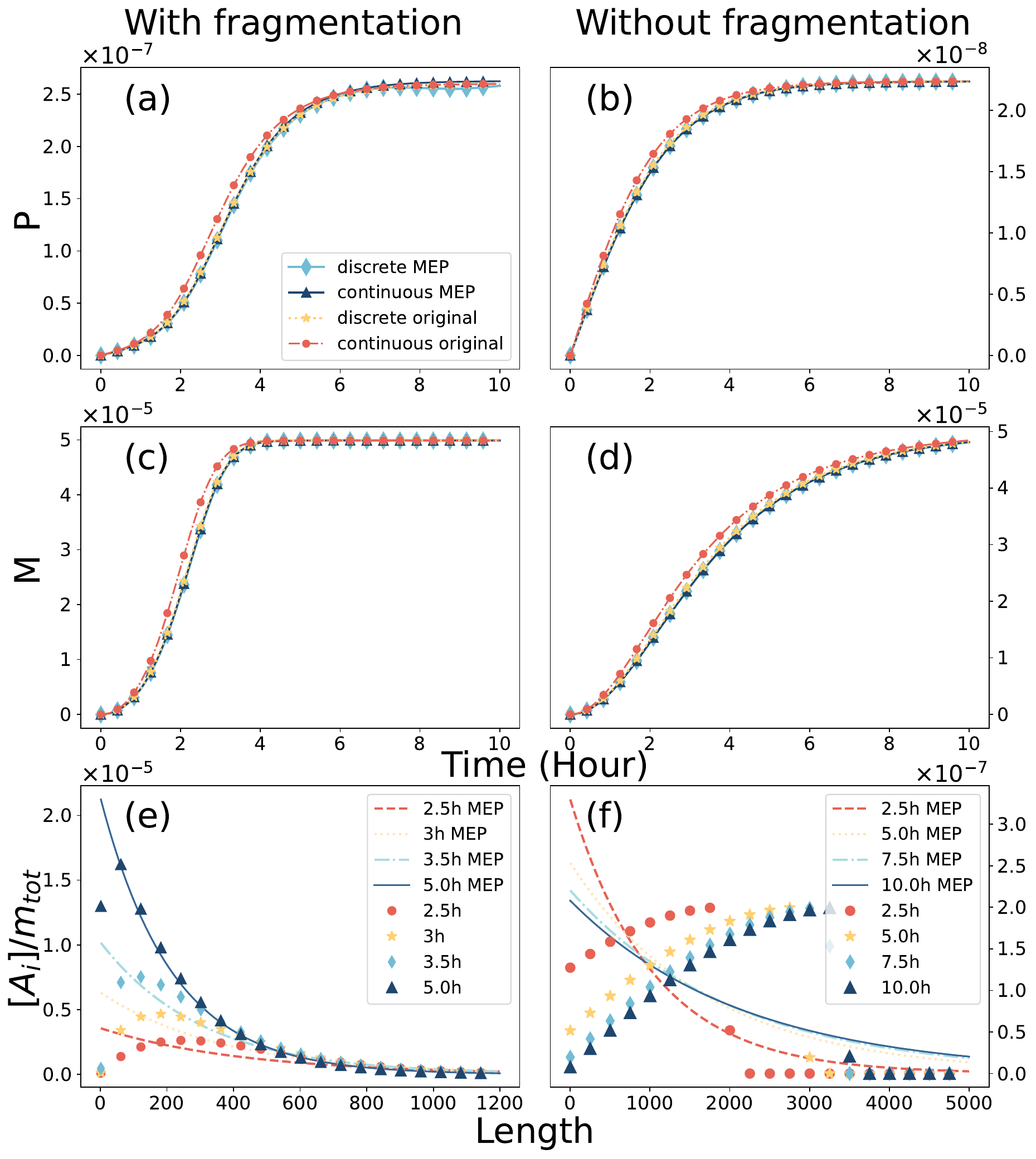}
    \caption{\textbf{Comparison between the full models and MEP-reduced models from the dynamical aspects.} Models with fragmentation are plotted in left panels (a,c,e), while those without fragmentation are given in right panels (b,d,f). Trajectories of  (a,b) the number concentration of fibrils $P$, (c,d) the mass concentration $M$, as well as (e,f) the normalized fibril length distribution $[A_i]/m_{tot}$ are highlighted here. For each graph on $P$ and $M$, four different curves presenting both continuous (continuous, continuous MEP) and discrete (discrete, discrete MEP) models are drawn separately. Additionally, fibril length distributions at four typical time points are compared. The parameters used for illustration are set as 
$n_c=2,k_n^+=1.0\times{10}^{-3}\mathcal{M}^{-1}s^{-1},k_n^-=7.3\times{10}^{-9}s^{-1},k_e^+=2.5\times{10}^{3}\mathcal{M}^{-1}s^{-1},k_e^-=2.5\times{10}^{-4}s^{-1},k_f^+=2.8\times{10}^{-6}s^{-1},k_f^-=2*\times{10}^{3}\mathcal{M}^{-1}s^{-1},m_{tot}=5.0\times{10}^{-5}\mathcal{\mathcal{M}}$.}
\label{Fig:kinetics}\end{figure}

Upon analyzing the chemical reactions in Eq. \eqref{CRN}, we find that the elongation processes do not affect the number concentration of fibrils $P(t)$, while the fragmentation has no influence on the mass concentration $M(t)$. This fact can be clearly observed from the reduced kinetic models in Eqs. \eqref{Eq:kineticseqdy} and \eqref{Eq:kineticseqcy}. 
Compared to the numerical solutions to Eqs. \eqref{dAdt} and \eqref{continuous model} given through dashed lines in Figs. \ref{Fig:kinetics}(a-d), the reduced models by the MEP indeed provide a good approximation on $P(t)$ and $M(t)$. Their finite difference can be attributed to the inconsistency in the fibril length distribution. As illustrated in Figs. \ref{Fig:kinetics}(e-f), the MEP assumes that the fibril length distribution adopts an exponential form, which does not fully agree with the explicit ones. 
However, in the presence of fragmentation processes, the difference in the fibril length distribution before and after simplification becomes negligible in the long-time limit (after 5 hours in our case); in contrast, if the fragmentation processes are not included, a dramatic distinction is observed, which reflects the limitations of the MEP. It is worth mentioning that even in this case solutions of the reduced models on $P(t)$ and $M(t)$ still match well with their corresponding ones based on the full models.

\section{Thermodynamics of the Reduced Models}
\label{sec:Results}
Now we are going to explore the thermodynamics of the reduced models for amyloid filaments formation based on the MEP. We present our results for
the discrete model with and without fragmentation, as well as their corresponding continuous limits one by one. 

\subsection{Discrete model without fragmentation}
\label{subsec:dy3}
By substituting the optimal length distribution of filaments $\{[A_i]\}_{i=1, i\geq n_c}$ in Eq. \eqref{optimal-dis} into Eq. \eqref{ent}, we arrive at an effective entropy function for the reduced model expressed through three macroscopic quantities $P, M$ and $m_{tot}$ as 
\begin{eqnarray}
\overline{Ent}\left(t\right)
=P[1-  \ln (PL_d^{-1})] + (m_{tot}-M)\left[ 1- \ln \left(m_{tot}-M\right)\right]-(M-n_cP)\ln (1-L_d^{-1}),
\label{Eq:dy3ent}
\end{eqnarray}
where $L_d =[M - (n_c - 1)P]/{P}$ stands for an effective average length. In the absence of fragmentation processes, the entropy production rate for the reduced model reads
\begin{equation}
\overline{epr}\left(t\right)=\left(\ln \hat{R}_n^+-\ln \hat{R}_n^-\right)\left( \hat{R}_n^+-\hat{R}_n^-\right)+2\left(\ln \hat{R}_e^+-\ln \hat{R}_e^-\right)\left( \hat{R}_e^+-\hat{R}_e^-\right)\geq 0.
\label{depr2}
\end{equation}
Each summand above represents the contributions from the primary nucleation and elongation, respectively.

Analogously, the free energy function for the reduced model becomes
\begin{equation}
\begin{array}{rl}
&\overline{F}(t)
=(M-M^e)-(P-P^e)+(M-n_cP)\left[\ln \left(1-L_d^{-1}\right)-\ln \left(1-(L_d^e)^{-1}\right)\right]\\
&+P\left[\ln PL_d^{-1}
-\ln P^e(L_d^e)^{-1}\right]+
\left(m_{tot}-M\right)\left[\ln \left(m_{tot}-M\right)-\ln \left(m_{tot}-M^e\right)\right],
\end{array}
\label{Eq:dy3F}
\end{equation}
in which \(P^e = \sum_{i=n_c}^{\infty}[A_i]^e\), 
\(M^e = \sum_{i=n_c}^{\infty}i[A_i]^e\) and $L_d^e =[{M^e-(n_c-1)P^e}]/{P^e}$ stand for values of \(P\), \(M\) and $L_d$ at the equilibrium. The corresponding free energy dissipation rate is 
\begin{equation}
\begin{array}{rl}
\overline{f_{d}}\left(t\right)
=\overline{epr}(t)
-\left(\ln \hat{R}_{n+}^e-\ln \hat{R}_{n-}^e\right)\left( \hat{R}_n^+-\hat{R}_n^-\right)-2\left(\ln \hat{R}_{e+}^e-\ln \hat{R}_{e-}^e\right)\left( \hat{R}_e^+-\hat{R}_e^-\right),
\end{array}
\label{Eq:dnfd}
\end{equation}
where $\hat{R}_{\alpha\pm}^e$ ($\alpha=n, e, f$) represent the equilibrium reaction rate functions by substituting the aforementioned equilibrium macroscopic physical quantities into Eq. \eqref{fluxofdiscrete}.
It is direct to verify that $\overline{f_{d}}\left(t\right)=\overline{epr}(t)$ once the condition of detailed balance holds.

\subsection{Discrete model with fragmentation}
\label{subsec:dn3}
By taking the fragmentation processes into consideration, the entropy function \eqref{Eq:dy3ent} and free energy function remain the same, but additional terms representing the contributions of fragmentation processes will come into play in both the entropy production rate and free energy dissipation rate, i.e.
By taking the fragmentation processes into consideration, the entropy function \eqref{Eq:dy3ent} and free energy function \eqref{Eq:dy3F} remain the same, but additional terms representing the contributions of fragmentation processes will come into play in both the entropy production rate and free energy dissipation rate, i.e.
\begin{eqnarray}
\label{depr3}
\widetilde{epr}\left(t\right)
&=&\overline{epr}(t)+\left(\ln \hat{R}_f^+-\ln \hat{R}_f^-\right)\left( \hat{R}_f^+-\hat{R}_f^-\right),\\
\widetilde{f_{d}}\left(t\right)
&=&\overline{f_{d}}\left(t\right)+\left(\ln \hat{R}_f^+-\ln \hat{R}_f^-\right)\left( \hat{R}_f^+-\hat{R}_f^-\right)
-\left(\ln \hat{R}_{f+}^e-\ln \hat{R}_{f-}^e\right)\left( \hat{R}_f^+-\hat{R}_f^-\right),
\label{dfd3}
\end{eqnarray}
both of which are non-negative. Similar to the model without fragmentation, the contributions from three different types of reactions are linearly combined together in these thermodynamic quantity.

Furthermore, provided the following condition of detailed balance holds, $\hat{R}_{\alpha+}^e=\hat{R}_{\alpha-}^e$, $\alpha=n, e, f$, the free energy dissipation rate equals to the entropy production rate, i.e. $\widetilde{f_{d}}\left(t\right)=\widetilde{epr}(t)$. Also, we observe that both Eqs. \eqref{depr2} and \eqref{depr3} are non-negative, and equal to 0 if and only if the system is at equilibrium, i.e. $P=P^e, M=M^e$.

As a summary, for the discrete model simplified by MEP, all thermodynamic quantities can be expressed in terms of the three moments \(P\), \(M\), and \(m_{tot}\), meaning that the thermodynamics of the reduced model presented above is purely macroscopic. 



The effective average length is an important quantity to characterize the formation process of amyloid filaments. Leave alone the initial lag-phase when only few nuclei are formed, the effective average length is usually much larger than unity. Thus, in the limit of $L_d\gg 1$, the above thermodynamic expressions could be cast into a much simpler form: 
\begin{eqnarray*}
\widetilde{Ent}\left(t\right)&\approx& P\ln (PL_d^{-1}) - (m_{tot}-M)\ln \left(m_{tot}-M\right),\\
\widetilde{F}(t)
&\approx&(M-M^e)-(P-P^e)+P\left[\ln PL_d^{-1}
-\ln P^e(L_d^e)^{-1}\right]\\
&&+
\left(m_{tot}-M\right)\left[\ln \left(m_{tot}-M\right)-\ln \left(m_{tot}-M^e\right)\right],\\
\widetilde{epr}\left(t\right)
&\approx&\left[\ln{k}_{n}^{+}\left(m_{tot}-M\right)^{{n_{c}}}-\ln k_n^-PL_d^{-1}\right]\left[{k}_{n}^{+}{\left(m_{tot}-M\right)}^{{n_{c}}}-{k}_{n}^{-}{PL_d^{-1}}\right]\\
&&+2\left[\ln {k}_{e}^{+}\left(m_{tot}-M\right)-\ln k_e^-\right]\left[{k}_{e}^{+}\left(m_{tot}-M\right)-{k}_{e}^{-}\right]P\\
&&+\left[\ln {k}_{f}^{+}-\ln k_f^-PL_d^{-1}\right]\left[{k}_{f}^{+}-{k}_{f}^{-}PL_d^{-1}\right]PL_d\geq 0,\\
\widetilde{f_{d}}\left(t\right)
&\approx&\widetilde{epr}(t)-\left[\ln{k}_{n}^{+}{\left(m_{tot}-M^e\right)}^{{n_{c}}}-\ln k_n^-{P^e(L_d^e)^{-1}}\right]\left[{k}_{n}^{+}{\left(m_{tot}-M\right)}^{{n_{c}}}-{k}_{n}^{-}{PL_d^{-1}}\right]\\
&&-2\left[\ln {k}_{e}^{+}\left(m_{tot}-M^e\right)-\ln k_e^-\right]\left[{k}_{e}^{+}\left(m_{tot}-M\right)-{k}_{e}^{-}\right]P\\
&&-\left[\ln {k}_{f}^{+}-\ln k_f^-P^e(L_d^e)^{-1}\right]\left[{k}_{f}^{+}-{k}_{f}^{-}PL_d^{-1}\right]PL_d\geq 0.
\end{eqnarray*}
Note that a further simplification by neglecting those terms accounting for backward reactions will lead to significant deviations.

\subsection{Continuous model without fragmentation}
\label{subsec:cy3}
In the continuous limit, the entropy function for the MEP-reduced model reads:
\begin{equation}
\overline{Ent}\left(t\right)
=P+\left(m_{tot}-M\right)-\left(m_{tot}-M\right)\ln \left(m_{tot}-M\right)-P\ln \left(PL_c^{-1}\right)+(M-n_cP)L_c^{-1},
\label{Eq:cy3ent}
\end{equation}
where $L_c =(M - n_c P)/{P}$. With respect to the above formula, the entropy production rate for the reduced model without the fragmentation processes is given by
\begin{equation}
\overline{epr}\left(t\right)=
\left(\ln \check{R}_n^+-\ln \check{R}_n^-\right)\left( \check{R}_n^+-\check{R}_n^-\right)+2\left(\ln \check{R}_e^+-\ln \check{R}_e^-\right)\left( \check{R}_e^+-\check{R}_e^-\right)\geq 0.
\end{equation}

In a similar way, the free energy function and its dissipation rate for the reduced model become 
\begin{eqnarray}
\overline{F}(t)
&=&(M-M^e)-(P-P^e)+(M-n_cP)\left[(L_c^e)^{-1}-L_c^{-1}\right]\nonumber\\
&&+P\left[\ln PL_c^{-1}
-\ln P^e(L_c^e)^{-1}\right]+
\left(m_{tot}-M\right)\left[\ln \left(m_{tot}-M\right)-\ln \left(m_{tot}-M^e\right)\right],\\
\label{Eq:cnfd}
\overline{f_{d}}\left(t\right)
&=&\overline{epr}(t)+2\left[L_c^{-1}-(L_c^e)^{-1}\right]\left(\check{R}_e^+-\check{R}_e^-\right)\nonumber\\
&&-\left(\ln \check{R}_{n+}^e-\ln \check{R}_{n-}^e\right)\left( \check{R}_n^+-\check{R}_n^-\right)-2\left(\ln \check{R}_{e+}^e-\ln \check{R}_{e-}^e\right)\left( \check{R}_e^+-\check{R}_e^-\right),
\end{eqnarray}
where \(P^e = \int_{i=n_c}^{\infty}[A_x]^edx\), 
\(M^e = \int_{i=n_c}^{\infty}x[A_x]^edx\) and $L_c^e =({M^e-n_cP^e})/{P^e}$ stand for values of \(P\), \(M\) and $L_c$ at the equilibrium state. $\check{R}_{\alpha\pm}^e$ ($\alpha=n, e$) represent the equilibrium reaction rate functions by substituting the aforementioned equilibrium macroscopic physical quantities into Eq. \eqref{fluxofcontinuous}.

\subsection{Continuous model with fragmentation}
\label{subsec:cn3}
To take the contribution of fragmentation processes into consideration, we only need to modify the entropy production rate and free energy dissipation rate as
\begin{eqnarray}
\widetilde{epr}\left(t\right)
&=&\overline{epr}\left(t\right)+\left(\ln \check{R}_f^+-\ln \check{R}_f^-\right)\left( \check{R}_f^+-\check{R}_f^-\right),\\
\widetilde{f_{d}}\left(t\right)
&=&\overline{f_d}(t)+\left(\ln \check{R}_f^+-\ln \check{R}_f^-\right)\left( \check{R}_f^+-\check{R}_f^-\right)
-\left(\ln \check{R}_{f+}^e-\ln \check{R}_{f-}^e\right)\left( \check{R}_f^+-\check{R}_f^-\right).
\end{eqnarray}

Remind that the continuation procedure makes sense only when the average length of filaments is much larger than one. This fact can be clearly seen from a direct comparison on formulas of free energy dissipation rate in the discrete and continuous reduced models. Compared to Eq. \eqref{Eq:dnfd}, a new remainder term $2\left[L_c^{-1}-(L_c^e)^{-1}\right]\left(\check{R}_e^+-\check{R}_e^-\right)$ appears in Eq. \eqref{Eq:cnfd}, which becomes negligible as long as $L_c \gg 1$ (see Fig. S3).


\begin{figure}[htbp]
    \centering
\includegraphics[width=1\linewidth]{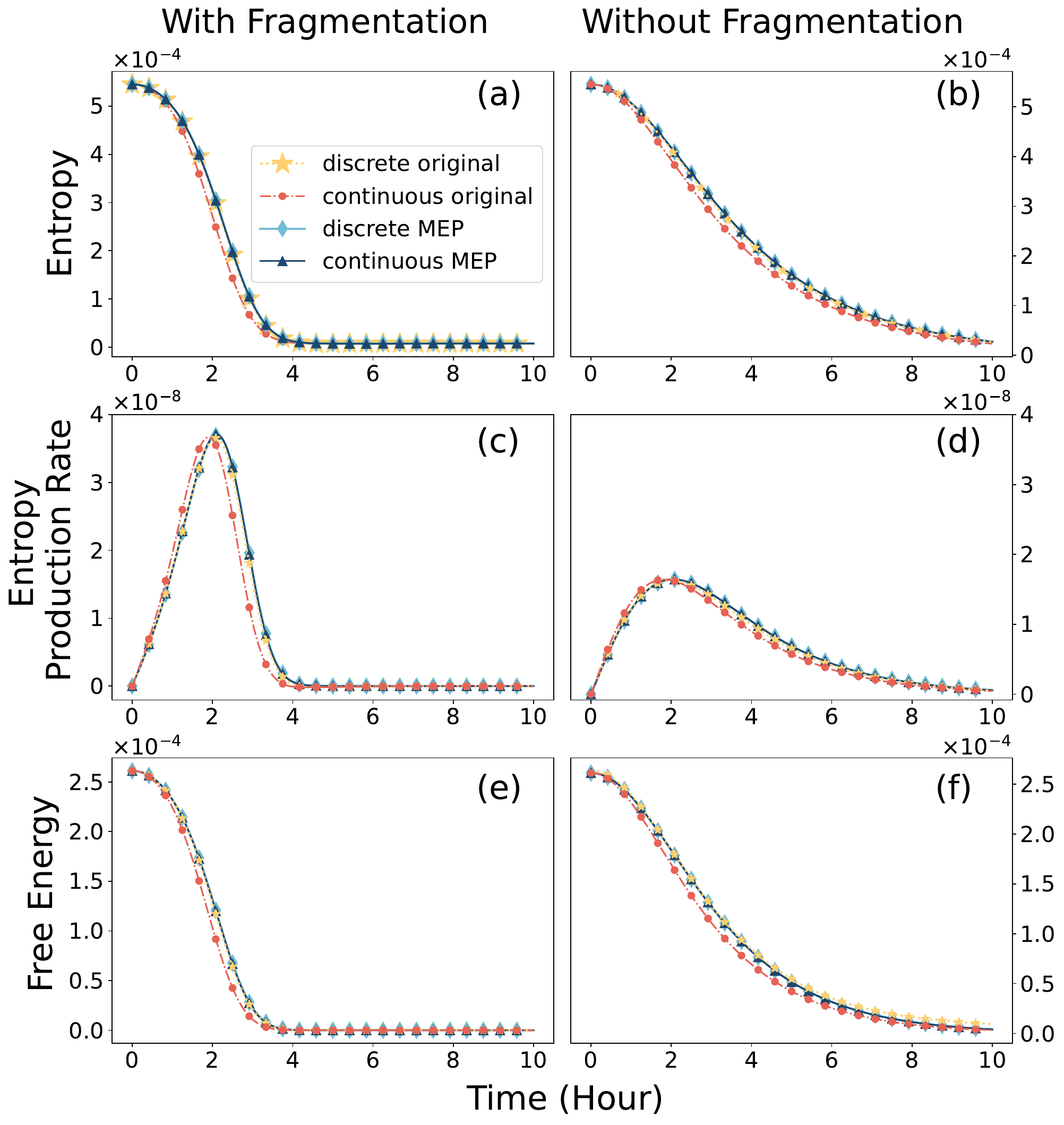}
    \caption{\textbf{Comparison on the thermodynamic quantities of the full models and MEP-reduced models.}
Models with (resp. without) fragmentation are plotted in left panels (a,c,e) (resp. right panels (b,d,f)). 
Each graph exhibits four curves presenting both continuous (continuous, continuous MEP) and discrete (discrete, discrete MEP) models separately.  Parameters are kept the same as those used in Fig. 2, which satisfies the condition of detailed balance.}
\label{Fig:Thermo}
\end{figure}

Above statements can be directly justified through numerical simulation. As shown in Fig. \ref{Fig:Thermo}, the thermodynamic quantities of the MEP-reduced models fit well with those of the original model. Given the parameters we adopted, which are chosen to satisfy the condition of detailed balance, the discrepancy between the continuous and discrete models are also quite small. Situations violating the detailed balance condition could be found Fig. S1 and S2 in the Supporting Information. At the same time, it is observed that the fragmentation processes make a significant contribution to the thermodynamic quantities. Not only the peak of the entropy production rate becomes approximately one half, but also the time evolution of other thermodynamic quantities, like the entropy and free energy, slow down in the absence of fragmentation reactions.

\section{Conclusion and discussion}
\label{sec:Conclusion}
In this study, through a concrete example about the formation kinetics of amyloid filaments, we demonstrate that the MEP provide a satisfactory way to do model reduction from both dynamical and thermodynamic aspects. The MEP-reduced models exhibit 
formally similar dynamical solutions, thermodynamic quantities and equilibrium states comparing to those of the original model. Due to this elegant consistency, we can establish the thermodynamics of the MEP-reduced models, which are expressed solely through macroscopic measurable quantities, for the first time.

During the theoretical derivations, we notice that the contributions of different types of chemical reactions are linearly additive in the kinetics equation of $P$ and $M$. Thank to this linearity, all quantities involving $\frac{dP}{dt}$ share a common feature. That is, the difference in the expressions of $\frac{dEnt}{dt}$, $J_f$, $epr$ and $f_d$ between models with fragmentation and without fragmentation is just an extra term related to fragmentation. In contrast, the expressions of those who do not involve $\frac{dP}{dt}$, like $\frac{dM}{dt}$, $Ent$, and $F$, are identical in the system with or without fragmentation.

Additionally, numerical simulations reveal that the non-equilibrium evolution processes of dynamical and thermodynamic quantities are significantly slower in the system lack of fragmentation reactions. This deceleration is particularly pronounced in the evolution of quantities involving $\frac{dP}{dt}$, such as $P$, $epr$, and $f_d$, leading to orders of magnitude differences observed in simulations. However, the overall trends of all quantities' evolution remain the same.

\section{CRediT authorship contribution statement}
\textbf{Xinyu Zhang}: Writing-review $\And$ editing, Writing-original draft, Visualization, Validation, Methodology, Formal analysis, Conceptualization. 

\textbf{Haiyang Jia}: Writing-review $\And$ editing, Validation, Methodology.

\textbf{Wuyue Yang}: Writing-review $\And$ editing, Validation, Methodology.

\textbf{Liangrong Peng}: Writing -
review $\And$ editing, Visualization, Validation, Formal analysis, Funding acquisition, Conceptualization. 

\textbf{Liu Hong}: Writing - review $\And$ editing, Visualization, Validation, Methodology, Project administration, Funding
acquisition, Conceptualization.
\section{Declaration of competing interest}
The authors declare that they have no known competing financial interests or personal relationships that could have appeared to influence the work reported in this paper.
\section{Acknowledgement}
This work was supported by the National Key R$\And$D Program of China (Grant No. 2023YFC2308702), the National Natural Science Foundation of China (12205135, 12301617), Guangdong Basic and Applied Basic Research Foundation (2023A1515010157).
\newpage

\bibliographystyle{ieeetr}
\bibliography{ref}

\begin{thebibliography}{10}

\bibitem{jaynes1985macroscopic}
E.~T. Jaynes, ``Macroscopic prediction,'' in {\em Complex Systems—Operational Approaches in Neurobiology, Physics, and Computers}, pp.~254--269, Springer, 1985.

\bibitem{Topsoe2007}
F.~Topsoe, ``Exponential families and maxent calculations for entropy measures of statistical physics,'' in {\em Complexity, Metastability and Nonextensivity} (S.~Abe, H.~Herrmann, P.~Quarati, A.~Rapisarda, and C.~Tsallis, eds.), vol.~965 of {\em AIP Conference Proceedings}, pp.~104--113, 2007.

\bibitem{kleidon2004non}
A.~Kleidon and R.~D. Lorenz, {\em Non-Equilibrium Thermodynamics and the Production of Entropy: Life, Earth, and Beyond}.
\newblock Springer Science \& Business Media, 2004.

\bibitem{harte2011maximum}
J.~Harte, {\em Maximum Entropy and Ecology: A Theory of Abundance, Distribution, and Energetics}.
\newblock OUP Oxford, 2011.

\bibitem{JMP2002}
J.~Ding and L.~R. Mead, ``{Maximum entropy approximation for Lyapunov exponents of chaotic maps},'' {\em Journal of Mathematical Physics}, vol.~43, no.~5, pp.~2518--2522, 2002.

\bibitem{JMP2018}
C.~Jin and J.~Ding, ``{A maximum entropy method for solving the boundary value problem of second order ordinary differential equations},'' {\em Journal of Mathematical Physics}, vol.~59, no.~10, p.~103505, 2018.

\bibitem{haarnoja2018soft}
T.~Haarnoja, A.~Zhou, P.~Abbeel, and S.~Levine, ``Soft actor-critic: Off-policy maximum entropy deep reinforcement learning with a stochastic actor,'' in {\em Proceedings of the 35th International Conference on Machine Learning} (J.~Dy and A.~Krause, eds.), vol.~80 of {\em Proceedings of Machine Learning Research}, pp.~1861--1870, PMLR, 2018.

\bibitem{Presse2013Principles}
S.~Press{\'e}, K.~Ghosh, J.~Lee, and K.~A. Dill, ``Principles of maximum entropy and maximum caliber in statistical physics,'' {\em Reviews of Modern Physics}, vol.~85, no.~3, p.~1115, 2013.

\bibitem{KenADill2012}
H.~Ge, S.~Presse, K.~Ghosh, and K.~A. Dill, ``Markov processes follow from the principle of maximum caliber,'' {\em Journal of Chemical Physics}, vol.~136, no.~6, 2012.

\bibitem{KenADill2019}
L.~Agozzino and K.~Dill, ``Minimal constraints for maximum caliber analysis of dissipative steady-state systems,'' {\em Physical Review E}, vol.~100, no.~1, 2019.

\bibitem{Anwasia2022}
B.~Anwasia and S.~Simic, ``Maximum entropy principle approach to a non-isothermal maxwell-stefan diffusion model,'' {\em Applied Mathematics Letters}, vol.~129, 2022.

\bibitem{Hong2013SimpleMM}
L.~Hong and W.~A. Yong, ``Simple moment-closure model for the self-assembly of breakable amyloid filaments,'' {\em Biophysical Journal}, vol.~104, no.~3, pp.~533--40, 2013.

\bibitem{Karlin2016}
I.~V. Karlin, ``Invariance principle and model reduction for the fokker-planck equation,'' {\em Philosophical Transactions of the Royal Society A-Mathematical Physical and Engineering Sciences}, vol.~374, no.~2080, 2016.

\bibitem{Horvat2015}
S.~Horvat, E.~Czabarka, and Z.~Toroczkai, ``Reducing degeneracy in maximum entropy models of networks,'' {\em Physical Review Letters}, vol.~114, no.~15, 2015.

\bibitem{JMP2022}
O.~Esen, M.~Grmela, and M.~Pavelka, ``{On the role of geometry in statistical mechanics and thermodynamics. I. Geometric perspective},'' {\em Journal of Mathematical Physics}, vol.~63, no.~12, p.~122902, 2022.

\bibitem{JMP2024}
O.~Esen, M.~Grmela, and M.~Pavelka, ``{On the role of geometry in statistical mechanics and thermodynamics. II. Thermodynamic perspective},'' {\em Journal of Mathematical Physics}, vol.~63, no.~12, p.~123305, 2022.

\bibitem{Hong2023}
L.~R. Peng and L.~Hong, ``Thermodynamics for reduced models of chemical reactions by pea and qssa,'' {\em Physical Review Research}, vol.~6, p.~013296, 2024.

\bibitem{Esposito2020}
F.~Avanzini, G.~Falasco, and M.~Esposito, ``Thermodynamics of non-elementary chemical reaction networks,'' {\em New Journal of Physics}, vol.~22, no.~9, 2020.

\bibitem{Chiti2006}
F.~Chiti and C.~M. Dobson, ``Protein misfolding, functional amyloid, and human disease,'' {\em Annual Review of Biochemistry}, vol.~75, pp.~333--366, 2006.

\bibitem{Molecules25051195}
Z.~L. Almeida and R.~M.~M. Brito, ``Structure and aggregation mechanisms in amyloids,'' {\em Molecules}, vol.~25, no.~5, 2020.

\bibitem{Louros2023}
N.~Louros, J.~Schymkowitz, and F.~Rousseau, ``Mechanisms and pathology of protein misfolding and aggregation,'' {\em Nature Reviews Molecular Cell Biology}, vol.~24, no.~12, pp.~912--933, 2023.

\bibitem{Spillantini1997}
M.~G. Spillantini, M.~L. Schmidt, V.~M. Lee, J.~Q. Trojanowski, R.~Jakes, and M.~Goedert, ``$\alpha$-synuclein in lewy bodies,'' {\em Nature}, vol.~388, no.~6645, pp.~839--840, 1997.

\bibitem{Selkoe2001}
D.~J. Selkoe, ``Alzheimer's disease: Genes, proteins, and therapy,'' {\em Physiological Reviews}, vol.~81, no.~2, pp.~741--766, 2001.

\bibitem{Aguzzi2010}
A.~Aguzzi and T.~O'Connor, ``Protein aggregation diseases: Pathogenicity and therapeutic perspectives,'' {\em Nature Reviews Drug Discovery}, vol.~9, no.~3, pp.~237--248, 2010.

\bibitem{Westermark2011}
P.~Westermark, A.~Andersson, and G.~T. Westermark, ``Islet amyloid polypeptide, islet amyloid, and diabetes mellitus,'' {\em Physiological Reviews}, vol.~91, no.~3, pp.~795--826, 2011.

\bibitem{LC2022}
D.~B. Kell, G.~J. Laubscher, and E.~Pretorius, ``A central role for amyloid fibrin microclots in long covid/pasc: origins and therapeutic implications,'' {\em Biochemical Journal}, vol.~479, no.~4, pp.~537--559, 2022.

\bibitem{OOSAWA196210}
F.~Oosawa and M.~Kasai, ``A theory of linear and helical aggregations of macromolecules,'' {\em Journal of Molecular Biology}, vol.~4, no.~1, pp.~10--21, 1962.

\bibitem{Lomakin1997}
A.~Lomakin, D.~S. Chung, G.~B. Benedek, D.~A. Kirschner, and D.~B. Teplow, ``On the nucleation and growth of amyloid beta-protein fibrils: Detection of nuclei and quantitation of rate constants,'' {\em Proceedings of the National Academy of Sciences}, vol.~93, no.~3, pp.~1125--1129, 1996.

\bibitem{Knowles2009Science}
T.~P.~J. Knowles, C.~A. Waudby, G.~L. Devlin, S.~I.~A. Cohen, A.~Aguzzi, M.~Vendruscolo, E.~M. Terentjev, M.~E. Welland, and C.~M. Dobson, ``An analytical solution to the kinetics of breakable filament assembly,'' {\em Science}, vol.~326, no.~5959, pp.~1533--1537, 2009.

\bibitem{Hong2017}
L.~Hong, C.~F. Lee, and Y.~J. Huang, ``Statistical mechanics and kinetics of amyloid fibrillation,'' in {\em Biophysics and Biochemistry of Protein Aggregation: Experimental and Theoretical Studies on Folding, Misfolding, and Self-assembly of Amyloidogenic Peptides}, pp.~113--186, 2017.

\bibitem{Knowles2015}
P.~Arosio, T.~P.~J. Knowles, and S.~Linse, ``On the lag phase in amyloid fibril formation,'' {\em Physical Chemistry Chemical Physics}, vol.~17, no.~12, pp.~7606--7618, 2015.

\bibitem{Dobson2014}
A.~K. Buell, C.~Galvagnion, R.~Gaspar, E.~Sparr, M.~Vendruscolo, T.~P.~J. Knowles, S.~Linse, and C.~M. Dobson, ``Solution conditions determine the relative importance of nucleation and growth processes in $\alpha$-synuclein aggregation,'' {\em Proceedings of the National Academy of Sciences}, vol.~111, no.~21, pp.~7671--7676, 2014.

\bibitem{deGroot1962}
S.~R. de~Groot and P.~Mazur, {\em Non-Equilibrium Thermodynamics}.
\newblock Amsterdam: North-Holland, 1962.

\bibitem{Hong2015}
L.~Hong, Y.~J. Huang, and W.~A. Yong, ``A kinetic model for cell damage caused by oligomer formation,'' {\em Biophysical Journal}, vol.~109, no.~7, pp.~1338--1346, 2015.

\bibitem{Sekimoto2010}
K.~Sekimoto, {\em Stochastic Energetics}.
\newblock Lecture Notes in Physics, Berlin, Heidelberg: Springer Berlin Heidelberg, 2010.

\bibitem{Sasa2022}
A.~Dechant, S.~Sasa, and S.~Ito, ``Geometric decomposition of entropy production into excess, housekeeping, and coupling parts,'' {\em Physical Review E}, vol.~106, no.~2, 2022.

\bibitem{Qian2017}
H.~Ge and H.~Qian, ``Mathematical formalism of nonequilibrium thermodynamics for nonlinear chemical reaction systems with general rate law,'' {\em Journal of Statistical Physics}, vol.~166, no.~1, pp.~190--209, 2017.

\end{thebibliography}

\end{document}